# NEW NOISE-BASED LOGIC REPRESENTATIONS TO AVOID SOME PROBLEMS WITH TIME COMPLEXITY

HE WEN[a, b], LASZLO B. KISH[a], ANDREAS KLAPPENECKER[c], AND FERDINAND PEPER[d]

[a]*Texas A&M University, Department of Electrical and Computer Engineering, College Station, TX 77843-3128, USA*

[b]*Hunan University, College of Electrical and Information Engineering, Changsha, 410082, China*

[c]*Texas A&M University, Department of Computer Science, College Station, TX 77843-33112, USA*

[d]*National Institute of Information and Communications Technology, Kobe, 651-2492 Japan*

**Abstract:** Instantaneous noise-based logic can avoid time-averaging, which implies significant potential for low-power parallel operations in beyond-Moore-law-chips. However, in its random-telegraph-wave representation, the complete uniform superposition (superposition of all *N*-bit binary numbers) will be zero with high probability, that is, non-zero with exponentially low probability, thus operations with the uniform superposition would require exponential time-complexity. To fix this deficiency, we modify the amplitudes of the signals of the L and H states and achieve an exponential speedup compared to the old situation. Another improvement concerns the identification of a single product-string (hyperspace vector). We introduce "time shifted noise-based logic", which is constructed by shifting each reference signal with a small time delay. This modification implies an exponential speedup of single hyperspace vector identification compared to the former case and it requires the same, $O(N)$ complexity as in quantum computing.

## 1. Introduction

*Noise-based logic* (*NBL*), a deterministic logic scheme, which is inspired by the fact that neural signals in the brain are stochastic processes, has recently been introduced [1-8] for low-power, large parallel operations in post-Moore-law-chips. In *NBL*, the logic information is carried by an orthogonal system of random noise processes, which forms the reference signal system (orthogonal base) of logic values [1]. Orthogonality of two noises $X_i$ and $X_j$ represent zero cross-correlation between them, that is, $\langle X_i X_j \rangle = 0$ for $i \neq j$, where the bracket represents time averaging procedure. To process the logic information anywhere in the noise-based data processor, access to the reference signal system is required. Generally, 2*N* independent (orthogonal) reference noises are needed to form *N* noise-bits, where in a single noise-bit *X*, one noise represents the *L* and another the *H* value of the bit, respectively. The *N*-bit long products (product-strings) of the base vectors are called *hyperspace vectors* because multiplication between noise-bits leads out from the original 2*N*-dimensional basis and form a hyperspace with $2^N$ dimensions [1, 2] . A logic signal that propagates in a single wire can be constructed by the binary superposition (on or off) of hyperspace vectors, which results in $2^{2^N}$ different logic values. Consequently, *N* noise-bits correspond to $2^N$ classical bits in a single wire [2]. If proper special-purpose operations on such a superposition could be executed with polynomial hardware and time complexity then *NBL* could be a potential competitor of quantum computation. To explore how neurones function in data processing, a new type of deterministic logic scheme has also been proposed [3] with a spike-based *NBL* and set-theoretical operations to form hyperspace and to utilize coincidence effects.

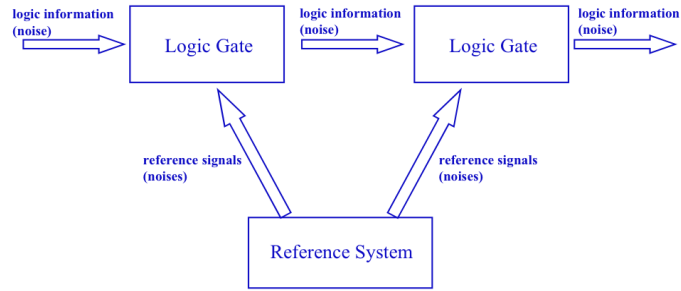

**Figure 1.** Generic outline of noise-based logic and computing

Recently, within the *NBL* framework, *instantaneous noise-based logic* (*INBL*) [4, 5] , has been introduced (note, the brain logic scheme [3] has also been identified as an *INBL* scheme) where the logic operations do not need to make statistics (evaluating correlations or time averages) on the noise-based signals. One possible realization of the *INBL* scheme is the solution where the logic values are encoded into nonzero, bipolar, independent *random-telegraph-waves* (*RTW*). The *RTW*s used in the *INBL* are random square waves, which take the value of +1 or -1 with probability 0.5 at the beginning of each clock period and stay with this value during the rest of the clock duration. In the non-squeezed *INBL*, for the *r-th* noise-bit, there are two independent reference *RTW*s, $H_r(t)$ and $L_r(t)$, representing its logic values, respectively.

|  | **Type of noise** | **Universal? (when Boolean)** | **Note** |
| --- | --- | --- | --- |
| **Correlator-based** | Arbitrary (continuum, random telegraph or spike waves, etc.) | Yes | Slow; error robust |
| **Instantaneous (squeezed and non-squeezed)** | Random telegraph waves<br><br>Random spike waves (brain) | Yes | Fast. Potential for large, parallel operations with low complexity. Error prone. |

Table 1. The most important noise-based logic schemes. Each of them is able to utilize the hyperspace

Although, the *INBL* has the advantages of avoiding time-averaging, there are still some problems, for example exponential complexity issues with the identification of hyperspace vectors or computations with the large superposition. In this Letter we propose solutions to fix these problems.

For the sake of notational simplicity, in the rest of this Letter, we omit the time variable (*t*) of stochastic time functions representing the noise-bits values.

**2. Some problems in *INBL***

*2.1 Zero value of the complete uniform superposition with high probability*

For applications mimicking quantum computing, the *N*-bit long hyperspace vectors (product-strings), $W_i = X_1 X_2 ... X_N$ and their superpositions have key importance [2]. These products correspond to the basis vectors of the $2^N$ dimensional Hilbert space of quantum informatics because there are $2^N$ orthogonal product-strings of length *N*. For example [2], with *N*=3, one has the 8 orthogonal hyperspace vectors

$$W_1 = L_1 L_2 L_3 , \quad W_2 = L_1 L_2 H_3 , \quad W_3 = L_1 H_2 L_3 , \quad W_4 = L_1 H_2 H_3 ,$$

$$W_5 = H_1 L_2 L_3 , \quad W_6 = H_1 L_2 H_3 , \quad W_7 = H_1 H_2 L_3 , \quad W_8 = H_1 H_2 H_3$$

(1)

The *complete* $2^N$ dimensional *uniform superposition,* which is the uniform superposition of all the *N*-bit long product-strings, and representing the superposition of all *N*-bit integer numbers, has complexity of $O(N*2^N)$. For applications, it is essential that the complete uniform superposition can be synthesized with low-complexity operations. With an Achilles-ankle operation it can be generated [2] with only $O(2N)$ complexity:

$$Y(2^N) = \prod_{r=1}^{N} (L_r + H_r) \tag{2}$$

The easy synthesis of the complete uniform superposition and the resulting high parallelism when manipulating it with simple, low-complexity operations, is very attractive. However, further analysis shows that the complete uniform superposition will be zero with a high probability. This is because each term in the product (2), i.e. ($L_r+H_r$), is non-zero with 0.5 probability, which yields the result that $Y(2^N)$ is non-zero with $0.5^N$ probability, only. Hence, the observation of non-zero values, which are essential for data processing and calculations, will require a waiting time exponentially long versus *N*.

*2.2 High time complexity for hyperspace vector identification*

Suppose, the result of calculations with *INBL* is a product-string $W_i$ described above, where $1 \leq i \leq 2^N$. To decode the result, we must identify $W_i$. In [6] a method was shown to decide if the $W_i$ is different from a *given* product-string. The required observation time for this operation was a logarithmic function of the probability that there is no answer during this observation time, which means about 83 clock steps for $10^{-25}$ or less chance of not getting an answer. Utilizing this method, to identify the $W_i$ from the set of the $2^N$ different product-strings*,* in the worst case, we must do the test with $2^N$-1 different product-string signals, which means on the average ($2^N$-1)/2 tests. In other words, the time complexity of the hyperspace vector identification is $O(2^N)$*,* which makes identifying product-string results difficult unless we must identify it from a small set of possible results.

On the other hand, in quantum computing, the complexity of the identification is only $O(N)$. To see this, we can assume a beam that consists of *N* photons of different colors, where each photon can be in one of the orthogonal polarization states representing the corresponding logic states *H* or *L*. Then assuming a diffraction device (e.g. prism) to send the photons of different colors to different directions, followed by *N* polarizing beam splitters, to send the photon of a given color but different polarization to different directions and 2*N* photon detectors placed at the outputs of the polarizing beam splitters we can identify this *N*-qubit beam with a single measurement provided the quantum efficiency is 1. Thus the complexity (hardware plus time) of this measurement is $O(N)$.

Thus, with the present form of *NBL* a quantum computer would need exponentially less complexity than *NBL* to decode the result. If this weakness of former *NBL* schemes cannot be fixed then quantum computing is superior to noise-based logic. Below we show the solution which guarantees that *NBL* performs this operation with the same $O(N)$ complexity as a quantum computer.

## 3. Modifying the reference base to improve computations with complete uniform superposition

Let us define the possible values of the *r*-th noise bit as follows. Suppose, $A_r$ and $B_r$ ($r = 1, ..., N$) be independent, discrete-time *RTW* signals with unity amplitude as earlier. The *r-th* noise-bit and its *High* and *Low* logic values are defined as

$$H_r = A_r \text{ and } L_r = \lambda B_r . \tag{3}$$

where λ (0<λ<1) is a constant number.

After the above modification, the complete uniform superposition in the new representation is

$$Y(2^N) = \prod_{r=1}^{N} \left( A_r + \lambda B_r \right) . \tag{4}$$

The particular advantage of this representation is that the absolute value of the time function $Y(2^N)$ of this superposition will never be zero.

As a practical example, let us consider λ=0.5 . Then the minimum and the maximum of the absolute value of the time function (4) are

$$Y(2^N)_{\min} = 0.5^N \text{ and } Y(2^N)_{\max} = 1.5^N , \tag{5}$$

respectively. To represent the whole range from $0.5^N$ to $1.5^N$, the required bit resolution *M* to emulate all possible superpositions of *N* -long products in a noise-based logic hardware is given as:

*Time-shifted noise-based logic*

$$M=N+N\log_2 1.5 < 2N \ . \tag{6}$$

It is important to remember that the complete uniform superposition is the superposition of $2^N$ binary numbers, where each number has $N$ bits resolution similarly to quantum computing. For example, when the number of noise-bits $N$=200, parallel operations on this superposition would require at least $200*2^{200}$ classical bits in a classical binary computer. Utilizing *INBL*, the $N$=200 noise-bits can be run on a classical physical noise-based hardware with only $M$=317 bits resolution. For example, in the *RTW*-based scheme described in Section 1, the NOT operation on the second bit of each of the $2^N$ numbers is simply done by multiplying the superposition with $H_2L_2$ [5] , which requires a hardware of $O(1)$ complexity. However, it is important to emphasize that such a trick can be made only for certain operations just like in quantum computing. The potential operations are subject of current research.

**4. Time-shifted noise-based logic to speed up hyperspace vector identification**

*4.1 Time-shifted noise-based logic*

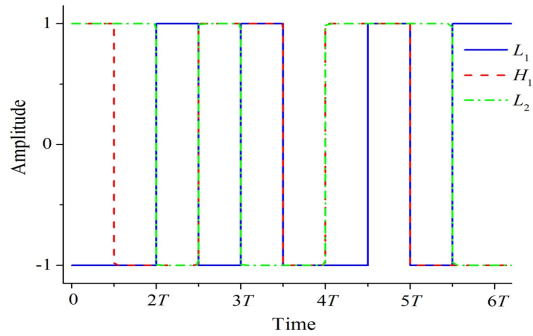

**Figure 2.** Example for the timing in the original *INBL* [4,5]. *RTW*s representing the logic values ($L_1$, $H_1$, $L_2$) of noise-bits are changing their amplitude values (signs) with 0.5 probability at the beginning of each clock period.

For the sake of simplicity but without limiting generality, in this section, we assume that the reference *RTW*s have unit amplitudes as in the original *INBL* scheme [4,5]. Originally, the reference signals that are representing the logic values (*High* and *Low*) of a single noise-bit are changing their states at the same time, namely, at the beginning of the clock period, see Figure 2. To improve the resolution of the *INBL* scheme, let us divide the clock period *T* into 2*N* sub-clock periods (*SCP*) where the duration of each *SCP* is τ=0.5*T/N*.

In the new, *time-shifted INBL* (*TSINBL*) scheme, the *RTW*s representing the logic values are shifted so that the initial time of each *SCP* coincides with the initial time of the corresponding *RTW*; i.e. each reference has a different moment of switching with 0.5 probability. In the example, shown in Figure 3, at *t*=0 the *RTW* representing $L_1$ steps; at *t*=τ the *RTW* representing $H_1$ switches; at *t*=2τ the *RTW* representing $L_2$ switches; etc.

It is important to note that the earlier hyperspace vector (product-string) is not lost and it can be read out in the last time segment of the clock signal because by then all the 2*N* reference signals reach their relevant value for the actual clock period.

The rest of the time segment, which requires a 2*N* fold increase of time resolution (complexity), serves only identification/separation purpose, that results in an exponential, $2^N$ fold speedup in the product-string identification, see below.

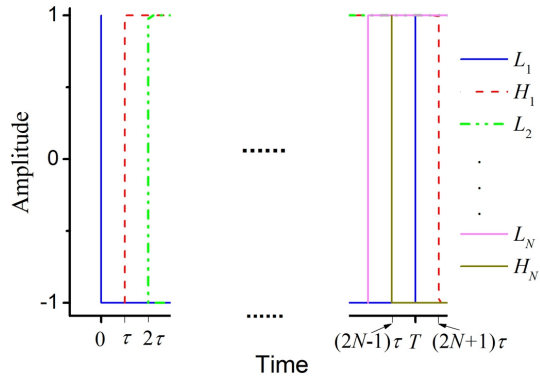

**Figure 3.** The initial time of each *SCP* coincides with the initial time of the corresponding *RTW* representing the different logic value of noise-bits in the *TSINBL*

### *4.2 Speedup by TSINBL in hyperspace vector identification*

Let us suppose we have an unknown product-string $W_i = \prod_{k=1}^{N}(X_k)$ of length *N*, where each *i* $(1 \leq i \leq 2^N)$ represents a different string (different *N*-bit long binary number). Each *RTW* representing its corresponding noise-bit value has a different start time in the *TSINBL* scheme. The necessary condition for making a decision about the question if a given noise-bit value is in the product or not, is that at least one of the two *RTW*s representing the noise-bit must change its state. At the particular *SCP* when a given reference *RTW* changes its state, the amplitudes of other reference *RTW*s remain constant. Consequently, if the product $W_i$ changes its amplitude at the same moment of time, we conclude that the given noise-bit value is in the string; while in the opposite situation its inverse value is there. For example, suppose that we detect a sign change of the amplitude $W_i$ and $L_3$ at the discrete time 4τ. Then we conclude that $W_i$ contains $L_3$ ($X_3 = L_3$). If only $L_3$ changes while $W_i$ remains the same the conclusion is that $H_3$ is the value of $X_3$.

The chance that the two *RTW* signals ($L_r$ and $H_r$) representing the possible values of the *r-th* noise-bit keep their sign during *M* clock periods is $0.25^M$. Therefore, in the limit of $0.25^M \leq\leq 1$ the probability of the "error" ε (*M*) of not obtaining a response at least for one of the noise-bits in the string over *M* clock periods, is exponentially small, ε (*M*)=$N*0.25^M$. Thus the time requirement of this operation scales with $2N*\log_4(1/ε)$. Hence, with this *TSINBL* scheme, the time complexity of the hyperspace vector identification is *O(N)*, which represents an exponential speedup compared to the classical *INBL* schemes [4-6].

## 5. Conclusions and Discussion

Although *INBL* has significant potential for low-power parallel operations without time-averaging, a number of open problems still exist. In this Letter, we have identified and proposed solutions for two problems: the problems of zero amplitudes of the complete uniform product-superposition and the exponential time requirement to identify an unknown product state.

**Acknowledgments**

This work has partially been supported by the National Natural Science Foundation of China under grant 61002035.